# Echo Planar Time-Resolved Imaging (EPTI) with Subspace Reconstruction and Optimized Spatiotemporal Encoding

Zijing Dong[1,2*], Fuyixue Wang[1,3], Timothy G. Reese[1], Berkin Bilgic[1,4], Kawin Setsompop[1,2,4]

[1]Athinoula A. Martinos Center for Biomedical Imaging, Massachusetts General Hospital, Charlestown, Massachusetts;

[2]Department of Electrical Engineering and Computer Science, MIT, Cambridge, Massachusetts;

[3]Harvard-MIT Health Sciences and Technology, MIT, Cambridge, Massachusetts;

[4]Department of Radiology, Harvard Medical School, Boston, Massachusetts.

___________________________________

Running Title: EPTI with subspace reconstruction and optimized encoding

Word Count: abstract (242), manuscript (~4900)

**\*Correspondence to:**

Zijing Dong,

Department of Electrical Engineering and Computer Science

Massachusetts Institute of Technology

Cambridge, MA, United States

Phone: 857-472-3722

Email: zijingd@mit.edu

**Grant sponsor:**

This work was supported by the NIH NIBIB (R01-EB020613, R01-EB019437, R01-MH116173, P41-EB015896, and U01-EB025162) and the instrumentation Grants (S10-RR023401, S10-RR023043, and S10-RR019307).

# Abstract

**Purpose:** To develop new encoding and reconstruction techniques for fast multi-contrast/quantitative imaging.

**Methods:** The recently proposed Echo Planar Time-resolved Imaging (EPTI) technique can achieve fast distortion- and blurring-free multi-contrast/quantitative imaging. In this work, a subspace reconstruction framework is developed to improve the reconstruction accuracy of EPTI at high encoding accelerations. The number of unknowns in the reconstruction is significantly reduced by modeling the temporal signal evolutions using low-rank subspace. As part of the proposed reconstruction approach, a $B_0$-update algorithm and a shot-to-shot $B_0$ variation correction method are developed to enable the reconstruction of high-resolution tissue phase images and to mitigate artifacts from shot-to-shot phase variations. Moreover, the EPTI concept is extended to 3D k-space for 3D GE-EPTI, where a new 'temporal-variant' of CAIPI encoding is proposed to further improve performance.

**Results:** The effectiveness of the proposed subspace reconstruction was demonstrated first in 2D GESE EPTI, where the reconstruction achieved higher accuracy when compared to conventional $B_0$-informed GRAPPA. For 3D GE-EPTI, a retrospective undersampling experiment demonstrates that the new temporal-variant CAIPI encoding can achieve up to 72× acceleration with close to 2× reduction in reconstruction error when compared to conventional spatiotemporal-CAIPI encoding. In a prospective undersampling experiment, high-quality whole-brain $T_2^*$ and QSM maps at 1 mm isotropic resolution was acquired in 52 seconds at 3T using 3D GE-EPTI with temporal-variant CAIPI encoding.

**Conclusion:** The proposed subspace reconstruction and optimized temporal-variant CAIPI encoding can further improve the performance of EPTI for fast quantitative mapping.

# Introduction

Echo planar imaging (EPI) (1) is a fast MR sequence that has been widely used for diffusion imaging (2) and functional imaging (3,4). However, the low bandwidth along the phase encoding (PE) direction of single-shot EPI (ss-EPI) can lead to severe image distortion and $T_2^*$ blurring, limiting EPI's ability to acquire high-quality images with accurate anatomical structures. To mitigate distortion and blurring, parallel imaging (PI) techniques (5,6) have been applied to ss-EPI by reducing its the effective echo spacing (ESP). However, the mitigation of distortion and blurring using PI is limited by the achievable in-plane acceleration. In the pursuit of higher distortion and blurring reduction, multi-shot EPI (ms-EPI) techniques (7-11) have been developed to further reduce the effective ESP at the cost of a longer acquisition time. Additionally, a navigator acquisition might be necessary to correct for the shot-to-shot $B_0$ variations due to respiration and physiological noises (12,13), which further increases the scan time.

Point-Spread-Function (PSF) mapping (14-17) is a unique ms-EPI technique that can achieve high-quality distortion- and blurring-free images, but requires the use of a very large number of acquisition shots. Recently, the tilted-CAIPI (18) acquisition/reconstruction scheme has been developed to accelerate PSF acquisition, which enables distortion- and blurring-free brain imaging at 1-mm resolution in ~8 EPI-shots per slice. PSF mapping with tilted-CAIPI has also been extended to diffusion imaging, where it was refined to enable self-navigation of shot-to-shot phase variation (18). To further improve the efficacy of tilted-CAIPI for multi-echo multi-contrast imaging, a new, related ms-EPI technique, termed Echo Planar Time-resolved Imaging (EPTI) (19,20) was developed. EPTI not only achieves distortion- and blurring-free imaging using a small number of EPI-shots, it also resolves hundreds of $T_2/T_2^*$-weighted images across the EPI readout at different echo times. The high acquisition efficiency and rich signal evolution information provided by EPTI has enabled fast multi-contrast and quantitative mapping. For example, whole-brain maps of proton-density, $T_2$, $T_2^*$, $B_0$, and susceptibility weighted imaging (SWI) at $1.1 \times 1.1 \times 3$ mm resolution can be acquired in under 30 seconds (19).

In recent years, a number of advanced reconstruction methods have been developed for accelerated quantitative mapping and multi-contrast imaging, including compressed sensing (21,22), low-rank priors (23-27), model-based approaches (28-30), and dictionary-based/pattern-match methods (31-33). In particular, the low-rank subspace methods (24,27,34) are able to resolve the temporal evolution of the signal across different contrasts with robustness to partial volume and multi-compartment effects, and have shown promising results for quantitative mapping. Additionally, the subspace can account for more complicated image artifacts and sequence imperfections such as $B_1$ inhomogeneity (24) through modeling the effects into the forward signal model. The good performance and flexibility of subspace-based reconstruction is leveraged in many applications including spectroscopic imaging (35-37), $T_2$-weighted knee imaging (24,38), quantitative brain (34,39) and cardiac imaging (40,41).

In conventional EPTI, a GRAPPA-like $B_0$-informed parallel imaging reconstruction is used to exploit the signal correlation along the temporal, spatial and coil dimensions to recover missing data-points in *k-t* space. In this study, we developed a subspace reconstruction framework tailored to EPTI to improve its reconstruction at very high accelerations. To accurately account for $B_0$-inhomogeneity phase accumulation across EPTI's multi-echo data, we incorporate a "$B_0$-update" algorithm (42) to recover both image magnitude and high-resolution $B_0$ information directly from the undersampled EPTI data. With this approach good reconstruction can be obtained without need for a priori high-resolution $B_0$ map that can be lengthy to acquire. Moreover, a data-driven 'shot-to-shot $B_0$-variation' correction method was also developed to provide robustness to shot-to-shot phase variation.

In addition to the outlined reconstruction developments, spatiotemporal CAIPI encoding of EPTI was also extended to 3D k-space for 3D GE-EPTI. A new 'temporal-variant' of CAIPI encoding was also developed and demonstrated to allows acceleration rates of up to 72× folds in *k-t* space. To maintain adequate image SNR for $T_2^*$ mapping at 3T, a more moderate acceleration factor of 32× was used in a prospectively accelerated acquisition, which enables whole-brain $T_2^*$ quantification at 1-mm isotropic resolution in just 52 seconds. The efficacy of this technique is verified with prospective and retrospective undersampling experiments.

## Theory

*Review of EPTI acquisition*

In EPTI, continuous EPI readouts are performed using highly-accelerated spatiotemporal CAIPI-sampling to efficiently sample the desired signal in *k-t* space. For instance, a 2D gradient- and spin-echo (GESE) EPTI sequence as shown in Fig. 1A can be used to acquire images across the $T_2/T_2^*$ signal decay. The EPTI encoding pattern along with a low-resolution fully-sampled calibration data in *k-t* space are shown in Fig. 1B. With EPTI encoding, neighboring readout lines are spaced apart in time by an ESP (Δt) and in PE direction by $R_{PE}$. Each shot acquires several diagonal $k_y$ line-sections that contain multiple readout lines with different PE encodings, with adjacent line-sections interleaved along PE to allow for better utilization of coil sensitivity. The sampling pattern is repeated temporally to ensure that the time distance between two $k_y$ line-sections ($T_s$) are regular and relatively short. This allows efficient use of temporal correlation during image reconstruction. $R_{seg}$ is the coverage along PE of each EPTI shot, which also determines the time distance between two adjacent $k_y$ line-sections, $T_s$, by $T_s = \Delta t \times R_{seg}/R_{PE}$. Shorter scan times can be achieved by using a larger $R_{seg}$, but this increases the time distance ($T_s$) leading to increased acceleration along *t*. Our previous work in Ref. 19 (19) has shown that for brain imaging at 3T with 1 mm in-plane resolution, 5-9 EPTI-shots with $R_{seg}$ of 20–40 and $T_s$ range of 5–10 ms can provide good image quality. As part of the acquisition, a fully-sampled low-resolution *k-t* calibration data is also acquired to train $B_0$-informed GRAPPA kernel for reconstruction (19).

*Subspace reconstruction framework*

*Subspace reconstruction*

EPTI acquires highly undersampled *k-t* dataset that contains hundreds of image contrasts that track the signal evolution. To resolve these images from the undersampled data, spatiotemporal correlation in *k-t* space should be efficiently used. The key concept of subspace reconstruction (27) is that the signal evolution space is low-rank, and can be represented by several low-rank subspace bases. Since the number of bases is much smaller than the number of contrasts or time points, the number of unknowns can be significantly reduced through estimating the coefficients of bases instead of multi-echo images. Such subspace approach exploits the temporal correlation of signals by reducing the temporal redundancy in the estimation process, with multi-channel coil sensitivity information also integrated to provide additional spatial encoding information for the reconstruction.

Figure 1C illustrates the proposed subspace reconstruction for EPTI. At first, the signal evolution space ($T_2/T_2^*$ decay for GESE-EPTI) within certain quantitative parameter ranges is simulated based on the Bloch equation and the acquisition parameters. Then, several basis vectors are extracted through principle component analysis (PCA) as $\phi_i$. These vectors form a low-dimensional subspace that can closely approximate the entire signal space of interest. Using these bases, the temporal image series can be calculated by $\phi c$, where $c$ is the coefficient maps of the bases. In this way, the degrees of freedom of reconstruction is reduced from the number of time points to the number of bases, improving the conditioning of reconstruction and the SNR of images. The subspace-constrained reconstruction is solved by,

$$\min_c \|UFSB\phi c - y\|_2^2 + \lambda R(c) \qquad (1)$$

where $B$ is the phase evolution across different image echoes due to $B_0$ inhomogeneity, $S$ is the coil sensitivity, $F$ is the Fourier transform operator, $U$ is the undersampling mask, and $y$ is the acquired undersampled EPTI data. The regularization term $R(c)$ can be incorporated to further improve the conditioning and SNR, and $\lambda$ is the control parameter of the regularization. After solving for $c$, the time-series of images can be recovered by $\phi c$.

*$B_0$-update algorithm*

The $B_0$ inhomogeneity-induced phase change and coil sensitivity can both be estimated from a low-resolution *k-t* calibration data as shown in Fig. 1B. However, low-resolution $B_0$ maps do not contain high-frequency information that if not incorporated into the reconstruction can cause image artifacts in areas with strong susceptibility. To avoid a time-consuming acquisition for a high-resolution $B_0$ map, a phase update algorithm named phase cycling (42) is incorporated into the proposed subspace reconstruction, to estimate the high-resolution $B_0$ map as shown in step 1 and 2 of Fig. 2. In step 1, the initial image is reconstructed by subspace reconstruction using the low-resolution $B_0$ map obtained from the fast calibration scan. In step 2, a high-resolution $B_0$ map is estimated by phase-cycling using the pre-reconstructed image and acquired data:

$$\min_{B_0} \|y - UFSIe^{j2\pi tB_0}\|_2^2 + \lambda g(B_0) \quad (2)$$

where $I$ is the multi-echo image magnitude with background phase estimated from the subspace reconstruction, $B_0$ is the high-resolution $B_0$ map to be estimated, and phase evolution can be calculated by $B = e^{j2\pi tB_0}$, where $t$ is the echo times. A wavelet constraint term $\lambda g(B_0)$ is incorporated to improve the conditioning of the $B_0$ estimation. The updated high-resolution $B_0$ map can then be used as input into the subspace reconstruction (step 1) to further improve the accuracy of magnitude images.

*Data-driven shot-to-shot $B_0$ variation correction*

Shot-to-shot $B_0$ variation can cause phase inconsistency in the EPTI data, which was shown to result in small local spatial smoothing in the reconstructed images (19). To mitigate this issue in (19), a data-driven approach was developed to estimate and correct for this $B_0$ variation that can work well with $B_0$-informed GRAPPA reconstruction. However, such approach is not compatible with the proposed subspace reconstruction and a new correction approach is developed here as outlined below.

Ref. 43 (43) proposed a navigator-free data-driven approach to estimate the mean $B_0$ change (0$^{th}$-order) in each receive coil of each acquisition shot ($B_{var}$), and correct for these variations in the raw k-space data to achieve improved reconstruction in the presence of shot-to-shot $B_0$ variation. In this work, a data-driven correction method that accounts for more spatial distribution of the $B_0$ variation is developed as an extension to this approach and incorporated into the proposed subspace reconstruction as shown in step 3 of Fig. 2. Here, the multi-echo EPTI data is used to fit the mean $B_0$ change of each shot to achieve improved accuracy than single echo estimation. A 2$^{nd}$-order polynomial $B_0$ variation map is estimated for each shot that best approximate the mean $B_0$ change across different coils ($B_{var}$) estimated using the method in (43).

Specifically, the mean $B_0$ change of each channel is first calculated as $B_{var}$ for each shot (43), and then the coefficients of the polynomial $B_0$ variation map $c_{var}$ are estimated by solving

$$\min_{c_{var}} \|MSI'Ac_{var} - B_{var}\|_2^2. \quad (3)$$

Here, $A$ is the 2$^{nd}$-order polynomial matrix, $I'$ represents the image series with phase obtained after step 1 and 2, $S$ is the coil sensitivity, and $M$ indicates the averaging operator. Using this approach, we can estimate the polynomial $B_0$ variation map $Ac_{var}$ of each EPTI shot, and correct the corresponding phase by $B = e^{j2\pi t(B_0 - Ac_{var})}$, to mitigate artifacts in the subspace reconstruction. The three reconstruction steps shown in Fig. 2 can be repeated several times to improve the accuracy of both the magnitude and phase of the reconstructed image time-series.

*3D spatiotemporal encoding*

The design of the spatiotemporal sampling in EPTI is important in allowing accurate signal recovery from highly undersampled data. With the proposed subspace reconstruction, the sampling pattern is no longer constrained to be fixed across $k_y$-t space as was in the case of $B_0$-informed GRAPPA reconstruction (to limit the number of GRAPPA kernels that needs to be trained). In order to design an optimized encoding for 3D EPTI, three types of spatiotemporal EPTI encodings were proposed and evaluated using a 3D GE-EPTI sequence: i) CAIPI-based sampling, ii) temporal-variant CAIPI sampling, and iii) block-wise random sampling.

Figure 3A shows the sequence diagram and an example of the encoding pattern of a 3D GE-EPTI acquisition. Following each excitation pulse, an EPTI readout is acquired with multiple readout lines shown as filled circles in the bottom-left $k_y$-$k_z$ plot. This forms a spatiotemporal CAIPI pattern within each $k_y$-$k_z$ block (with $k_y$ samples interleaving across time) that is acquired during each EPTI readout. Here, for the purpose of illustration, several echoes (6 echoes in Fig. 3A) are grouped together shown as an 'echo-section' even if they are acquired at different TEs. The undersampling factor in $k_y$-$k_z$-t space is then determined by the block size, R = $k_y$-blocksize × $k_z$-blocksize, since only one readout line will be acquired at each TE within the block. For example, the block size of Fig. 3A is 12 × 6 ($k_y$-$k_z$) and the acceleration factor is 72. In each TR, the EPTI readout block covers a portion of the 3D k-space (blue or red boxes in Fig. 3A bottom-right), and after hundreds of TRs, the full $k_y$-$k_z$-t space will be covered.

The details of the three different *k-t* sampling trajectories are shown in Fig. 3B. The CAIPI-based sampling (top row) repeats the same block-wise spatiotemporal encoding pattern in all echo-sections, whereas the temporal-variant CAIPI sampling (middle row) shifts all the even echo-sections along $k_y$ and $k_z$ by certain amounts to achieve a more complementary sampling with the odd echo-sections. The block-wise random sampling (bottom row) on the other hand randomly samples a $k_y$-$k_z$ point within the block at each echo time. The block-wise encoding design used in all these three trajectories was chosen to ensure that the PE and partition gradient blips are small enough to avoid increasing ESP by a large amount and/or causing strong nerve stimulation during the acquisition.

Additionally, two variable density sampling (VDS) patterns: i) VDS temporal-variant CAIPI and ii) VDS block-wise random sampling, were also designed and utilized with locally-low rank (LLR) constraint to further improve the image SNR. The evaluation of different encoding patterns helps us investigate the factors that will affect the reconstruction performance of 3D EPTI, such as complementary $k_y$-$k_z$ pattern across time and uniform or non-uniform k-space sampling pattern.

## Methods

All data were acquired with a consented institutionally approved protocol on a Siemens Prisma 3T scanner with a 32-channel head coil (Siemens Healthineers, Erlangen, Germany).

*Evaluation of reconstruction using 2D GESE EPTI*

*Subspace reconstruction*

A fully-sampled 2D GESE EPTI dataset was acquired to evaluate the subspace reconstruction with the following acquisition parameters: FOV = 240 × 240 mm$^2$, in-plane resolution = 1.1×1.1 mm$^2$, slice thickness = 3 mm, number of shots = 216, number of echoes (GE/SE) = 40/80, ESP = 0.93 ms, echo time range of GE / SE = 8.4–44.7 ms / 70.8–144.3 ms, TR = 2.5 s. This data was then retrospectively undersampled in $k_y$-$t$ space by R$_{seg}$ = 24 to synthesize a 9-shot 2D EPTI acquisition. Subspace reconstruction and $B_0$-informed GRAPPA reconstruction were used to reconstruct the 9-shot data, and the results compared with the 216-shot fully-sampled reference.

The subspace bases were generated using the Bloch equation and a large range of quantitative tissue parameters (44): range of T$_2$ = 1–600 ms, range of T$_2^*$ = 1–500 ms. To simulate the effect of imperfect refocusing pulse due to B1 inhomogeneity that will change the magnitude ratio between SE and GE signals, a scaling factor of 0.8-1.2 between GE and SE was also included. Eight bases were extracted from the simulated signals by PCA, which can approximate the simulated signal space with an error < 0.2%. The stop criterion of the subspace reconstruction was a maximum iteration = 60.

*$B_0$-update evaluation*

The effectiveness of the $B_0$ update algorithm was evaluated using the same 2D GESE EPTI dataset. The high-resolution $B_0$ map and tissue phase were calculated using the fully-sampled data with a matrix size of 216 × 216 × 40 ($k_x$-$k_y$-$N_{echo}$) as reference, while a low-resolution calibration data with a matrix size = 49 × 216 × 6 was used to estimate the initial low-resolution $B_0$ map. The reconstructed magnitude and phase images from the subspace reconstruction with and without $B_0$ update were evaluated for the retrospectively undersampled 9-shot EPTI case. For the $B_0$ estimation, wavelet constraint was used with a lambda = 0.02 to improve the conditioning. The number of iteration was set to 50 for the $B_0$ estimation, and 5 for the large loop of subspace reconstruction and $B_0$ estimation.

*Data-driven $B_0$-variations correction*

To evaluate the proposed shot-to-shot $B_0$-variation correction method, a 9-shot 2D GE-EPTI data were simulated with added $B_0$-variation in the same way as in the evaluation performed in (19). Realistic $B_0$ maps obtained from an *in vivo* scan with a spin and gradient echo EPI (SAGE-EPI) sequence (45) were used, which contain $B_0$ variations across the 9 shots of up to 3 Hz as shown in Fig. 6A. The error maps of the subspace reconstruction with and without $B_0$-variation correction were calculated at different echo times and compared. Additionally, PSF analysis was performed to quantify the mitigated local smoothing achieved with $B_0$-variations correction. The PSF of 3D GE-EPTI acquisition was also analyzed with the same level of $B_0$-variation (standard deviation = 0–2 Hz) as in 2D EPTI, to demonstrate that the effect of local smoothing in 3D acquisition is negligible and therefore the shot-to-shot $B_0$-variation correction was only used for 2D EPTI acquisition.

### 3D GE-EPTI with spatiotemporal encoding

*Retrospective undersampling experiment*

Fully-sampled 3D GE-EPTI data was acquired and used to evaluate the performance of different spatiotemporal encoding patterns. The acquisition parameters were: FOV = 220 × 220 × 106 mm$^2$, resolution = 1.1 × 1.1 × 1.1 mm$^2$, matrix size = 200 × 200 × 96, number of shots = 200 × 96 (*y*-block × *z*-block) = 19200, TR = 80 ms. The total acquisition time was 25 minutes and 36 seconds. 50 echoes were acquired in each TR with an ESP of 0.93 ms, covering a TE range of 9.1 ms–54.7 ms. In the first experiment, the data was retrospectively undersampled in $k_y$-$k_z$-$t$ space by 72× folds with a block size of 12×6 ($k_y$-$k_z$) for all the three encoding strategies as shown in Fig. 3B, which reduces the acquisition time from >25 minutes to 22 seconds. The reconstruction errors of the three different undersamplings were calculated using the fully-sampled data as the reference. Moreover, VDS temporal-variant CAIPI and VDS block-wise random sampling were also evaluated at the same net acceleration factor of 72×, with locally-low rank (LLR) constraint utilized in the reconstruction. The VDS patterns are using elliptical sampling (no sampling outside the ellipse in $k_y$-$k_z$), and have two different sampling rates in the ellipse: 32× undersampling at the k-space center, 72× undersampling at the outer k-space.

For temporal-variant CAIPI sampling, reconstruction errors for acquisitions with different $k_y$ and $k_z$ shifts between odd and even echo-sections were compared to investigate the key factors in encoding design that can effect performance. Here, more vs. less complementary $k_y$-$k_z$ patterns between the odd and even echo-sections was analyzed. Using the optimized VDS temporal-variant CAIPI sampling, quantitative T$_2^*$ maps and tissue phase maps were calculated at undersampling factors of 72 (block size = 12×6) and 32 (block size = 8×4). Error maps were calculated using the fully-sampled data as reference.

*Prospective undersampling experiment*

Prospective undersampled 3D GE-EPTI data was acquired to demonstrate the usage of optimized 3D EPTI for fast *in vivo* quantitative mapping at high spatial resolution. The data was acquired with VDS temporal-variant CAIPI sampling at an acceleration factor of 32 (block size = 8×4). Other acquisition parameters were: FOV = 216 × 216 × 96 mm$^2$, resolution = 1 × 1 × 1 mm$^2$, matrix size = 216 × 216 × 96, ESP = 0.95 ms, number of echoes = 57, TE range = 3.8 ms–57 ms, number of shots = 27 × 24 (*y*-block × *z*-block) = 648, flip angle = 20°, TR = 80 ms. The total acquisition time was 52 s. In addition to the imaging data acquisition, a low-resolution calibration data was also acquired with the same FOV and ESP. The acquisition time of calibration data was ~18 s, where accelerations along $k_y$ and $k_z$ were 2 × 2 and only 8 echoes were acquired to reduce the acquisition time (TR = 30 ms). The key acquisition parameters were: matrix size = 216 × 42 × 32, the k-space center was fully-sampled to provide GRAPPA calibration, and the net acceleration factors of $k_y$ and $k_z$ = 1.6 and 1.5 respectively.

**Results**

Figure 4 shows the comparison of subspace and $B_0$-informed GRAPPA reconstructions for the 2D GESE dataset. In Fig. 4A, the reconstructed images and error maps at four different TEs are presented.

Both methods performed well for the center echoes of the EPTI readout window (TE = 27 and 107 ms), however, the subspace method significantly reduces the reconstruction errors of the edge echoes (TE = 14 and 141 ms) where the k-t GRAPPA kernel cannot be applied fully to fill in the missing data. Such improvement can also be observed in the reconstructed signal evolution across GE-SE signals of a representative voxel in the image as shown in Fig. 4B, where the subspace reconstruction recovers more accurate signal curve with reduced noise and errors, especially at the edge echoes.

The results from the $B_0$-update evaluation are shown Figure 5. The first row shows the high-resolution $B_0$ map and tissue phase calculated from the fully sampled calibration dataset with 216 PE and 40 echoes, which took 120 seconds to acquire for 10 slices. The second row shows that a low-resolution calibration dataset with 49 PE × 6 echoes can reduce the acquisition time to just 10 seconds, but the high-resolution tissue phase information is missing in this data, causing artifacts in the reconstructed image as can be seen in the error map. Using the $B_0$-update algorithm, the high-resolution tissue phase can be well estimated, and the image artifacts mitigated with a reduced root-mean square error (RMSE) as shown in the bottom row of the figure.

Figure 6B shows the evaluation results of the proposed $B_0$-variation correction, where the reconstructed images and the corresponding error maps are shown for reconstructions with and without $B_0$ variation correction. Figure 6C shows the standard deviation of the $B_0$-variation across the 9 simulated EPTI-shots, before and after correction. Without the correction, high standard deviations in the $B_0$ variation can be observed in the anterior portion of the brain, causing noticeable reconstruction errors in this area, particularly at long TEs where the phase accrual from such variation is more pronounced. With the proposed correction, the standard deviations is markedly reduced and this resulted in less image errors with smaller RMSEs. The effects of $B_0$-variation correction on the PSF are shown in Supporting Figure S1 at the same TE of 70 ms. In the 2D acquisition case, a $B_0$ variation with a standard deviation of 1.5 Hz causes local smoothing with a side lobe level of 4.1%, while after correction, the side lobe is reduced to only 2.1%. Supporting Figure S1B shows the PSF analysis for 3D EPTI where the smoothing effect of $B_0$ variation is spread along both $z$ and $y$ direction. In this case, the largest side lobe is only 0.8%, even without any correction, causing negligible smoothing.

Figure 7 shows the results of the retrospective undersampling experiment for 3D GE-EPTI at 72× acceleration. Five different encoding patterns were compared: CAIPI-based sampling, temporal-variant CAIPI sampling, block-wise random sampling, VDS temporal-variant CAIPI, and VDS random sampling. The $k_y$-$k_z$ views of these patterns in Figure 7 shows the summation of all the $k$-$t$ samples along the time dimension. The first three encodings were reconstructed using the proposed subspace approach without LLR constraint. Here, the temporal-variant CAIPI shows the best performance with the lowest RMSE (6.94%) when compared against CAIPI-based sampling (11.4%) and block-wise random sampling (8.56%). The last two encoding patterns which utilize VDS were reconstructed with additional LLR constraint. Here, the reconstruction from VDS temporal-variant CAIPI shows a lower error (5.89%) when compared against VDS random sampling (6.97%). Based on these results, the temporal-variant CAIPI sampling achieved better performance than CAIPI-based sampling and random sampling, and when combing with VDS and LLR, the temporal-variant CAIPI still shows better results than random sampling.

To further investigate the optimal sampling design based on the temporal-variant CAIPI scheme, reconstructions from data with different $k_y$ and $k_z$ shift steps between the odd and the even echo-sections were analyzed. Figure 8A shows the reconstruction RMSEs at different shift steps for an acquisition block size of 12×6 ($k_y \times k_z$). Here, the encoding pattern with 0-0 ($k_y$-$k_z$) shift corresponds to conventional spatiotemporal CAIPI-sampling, which has the largest error. The sampling patterns in $k_y$-$k_z$ view are shown in Fig. 8B&C for two representative shifting patterns: 4-1 shift (Pattern 1) and 12-2 shift (equivalent to 0-2 shift, Pattern 2). The 'total pattern' in $k_y$-$k_z$ (summation of the two echo-sections along $t$) of Pattern 2 contains more complementary sampling than Pattern 1, and achieved much lower reconstruction RMSE, indicating that a more complementary sampling can improve reconstruction.

The evaluation of quantitative $T_2^*$ and tissue phase maps obtained from the retrospective undersampling experiment is presented in Fig.9, at two undersampling factors of 72× and 32×, both with VDS temporal-variant sampling. The scan time of the fully-sampled data was 25 minutes at 1.1 mm isotropic resolution, which is reduced to 21 s and 48 s through 72× and 32× accelerations, respectively. The quality of the tissue phase obtained from the 72× accelerated EPTI is comparable to the full-sampled case with a small increase in the noise level, while the $T_2^*$ map shows higher error at higher noise level. With the 32× accelerated EPTI, the $T_2^*$ map has higher SNR and reduced error, while the acquisition time is still less than 1 minute.

The prospective undersampling results are shown in Fig. 10, where 3D GE-EPTI with VDS temporal-variant sampling was used to acquire whole-brain $T_2^*$-weighted images with 57 echoes at 1 mm isotropic resolution in 52 seconds. High-quality multi-echo $T_2^*$-weighted images, along with quantitative $T_2^*$ and tissue phase maps are presented in three orthogonal views.

**Discussion and Conclusion**

A subspace reconstruction framework for EPTI was successfully developed and demonstrated to provide improved reconstruction accuracy over conventional $B_0$-informed GRAPPA. Such reconstruction uses prior low rank information on the signal model to reduce the number of unknowns and improve the image SNR. In addition to taking advantage of low rank modeling, approaches for $B_0$ estimation and shot-to-shot $B_0$-variation correction were developed and demonstrated to further improve reconstruction and enable the use of fast low-resolution calibration scan. Moreover, the EPTI concept was extended successfully to 3D sampling and new 3D spatiotemporal CAIPI encoding schemes were developed to improve acceleration capability.

Through volumetric encoding and improved noise averaging, 3D EPTI can achieve higher SNR than their 2D EPTI counterpart, lending itself well for the application in rapid high-resolution multi-contrast and quantitative mapping. The additional partition encoding dimension in 3D EPTI provides more flexibility for undersampling, but also adds more complexity for the design of the spatiotemporal sampling pattern. In this work, various 3D spatiotemporal sampling patterns were compared using the subspace reconstruction. Among all of these strategies, the temporal-variant CAIPI pattern achieves the best performance (Fig. 7), and its combination with VDS and LLR constraint was demonstrated to further improve reconstruction accuracy at high accelerations.

The analysis of the encoding patterns in Fig. 8 shows that a more complementary sampling between different echo-sections can improve the reconstruction accuracy. The improvement reflects the benefits of creating more spatial and temporal correlation across the sampled $k$-$t$ signals. Specifically, a more complementary sampling in $k_y$-$k_z$ space will create more spatial correlation and allow for better signal recovery using coil sensitivity information and parallel imaging. Therefore, undersampling factor in $k_y$-$k_z$ is chosen to achieve high temporal correlation while keeping within the capability of parallel imaging reconstruction. The proposed temporal-variant CAIPI encoding achieved the best performance by creating such spatiotemporal correlation in the sampled data, and allows high undersampling of up to 72×.

The high acceleration and the high acquisition efficiency of EPTI with continuous readout enables ultra-fast multi-contrast and quantitative mapping at high isotropic resolution in human brain, as demonstrated by both retrospective (Fig. 9) and prospective (Fig. 10) undersampling experiments. Interestingly, better reconstruction performance with lower noise was achieved in the prospective undersampling case (Fig. 10) when compared to the retrospective case (Fig. 9), both at 32× acceleration, even though the latter was performed at a larger voxel size (1.1 mm vs. 1mm isotropic). This is likely to be because of the increased physiological noise and the inevitable motion during the lengthy acquisition of the fully sampled dataset (~25 minutes) used in the retrospective undersampling case. In an even higher acceleration case of 72×, the prospective sampling experiment was able to achieve good reconstruction of the magnitude and phase images. However, the magnitude images were somewhat noisy, which limits the ability to use them to fit for a high-quality quantitative $T_2^*$ map. Application of 3D EPTI at higher field strength, such as at 7T, should help boost the SNR and enable higher performance at such a high acceleration rate.

The combined use of temporal-variant CAIPI sampling and the proposed subspace reconstruction framework was demonstrated to provide high quality imaging at very high acceleration factors for 3D EPTI. These sampling and reconstruction approaches should also be useful for the recently developed inversion-prep 3D EPTI scheme for rapid quantitative mapping of $T_1$ $T_2$ and $T_2^*$ (46), and for the propeller-based EPTI (PEPTIDE) approach that can achieve motion-robust and fast quantitative imaging (47). Future developments along these directions as well as in the incorporation of the low-rank tensor modeling and multi-compartment analysis should open up new exciting opportunities in creating rich, multi-dimensional time-resolved imaging data rapidly and robustly.

**Acknowledgements**

This work was supported by NIH NIBIB (R01-EB020613, R01-EB019437, R01-MH116173, P41-EB015896 and U01-EB025162) and by the MGH/HST Athinoula A. Martinos Center for Biomedical Imaging; and was made possible by the resources provided by NIH Shared Instrumentation Grants S10-RR023401, S10-RR023043, and S10-RR019307.

**Figures**

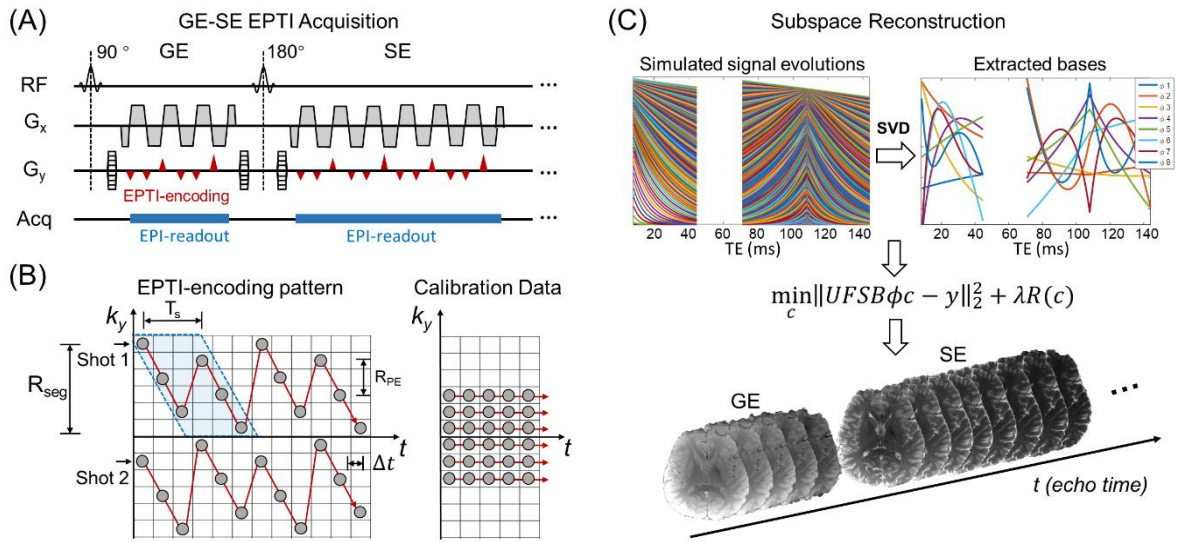

**FIG. 1.** The graphic illustration of EPTI acquisition and subspace reconstruction. (A) The diagram of 2D GE-EPTI sequence, and (B) an example of EPTI encoding pattern and fully-sampled calibration in $k_y$-$t$ space. In EPTI encoding, two adjacent $k_y$ line-sections are interleaved in the PE direction (as show in the blue block), and separated in time by $T_s$. $\Delta t$ is the ESP, $R_{PE}$ is the undersampling factor of two sequentially PE lines, and $R_{seg}$ is the segment size along PE direction of each shot. Figure 1C shows that in subspace reconstruction, signal evolution curves with $T_2/T_2^*$ decay are simulated based on the tissue and acquisition parameters. The subspace bases are extracted by PCA, and used in reconstruction to approximate the actual signal temporal evolution.

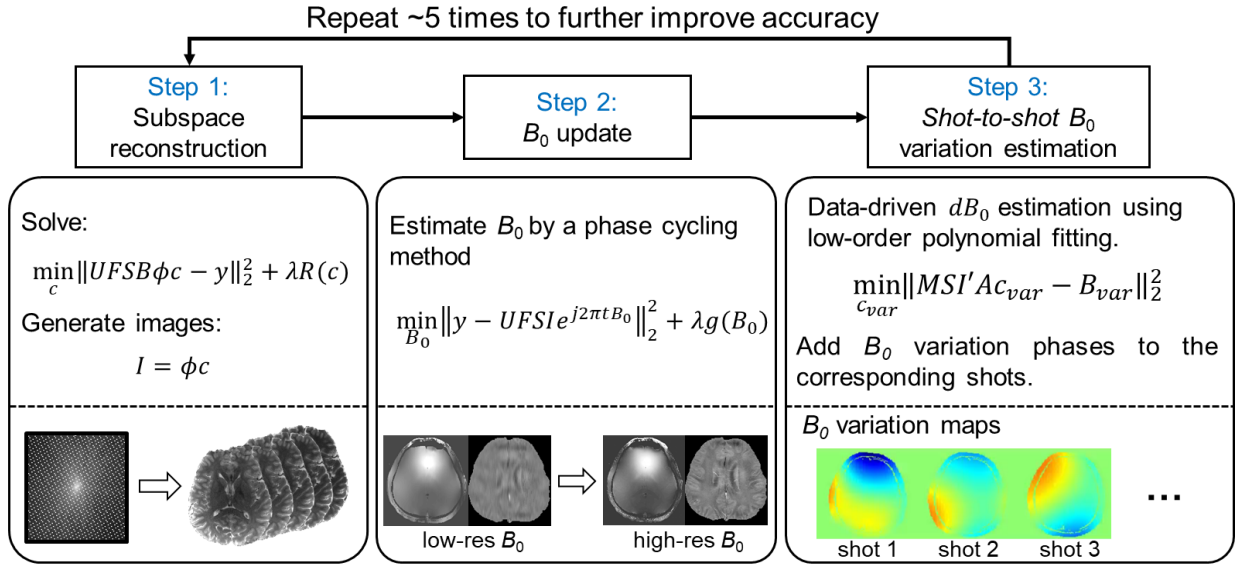

**FIG. 2.** Flowchart of the subspace reconstruction with $B_0$ update and shot-to-shot $B_0$ variation correction. In step 1, image magnitude is estimated by subspace reconstruction using the low-resolution $B_0$ map from the fast calibration data. In step 2, the high-resolution phase can be estimated by the $B_0$ update algorithm with fixing image magnitude. In step 3, shot-to-shot phase variation of each shot can be estimated from the undersampled data, and applied to the phase evolution in the reconstruction model. The large loop of the three steps can repeat several times to improve the accuracy of both image magnitude and phase.

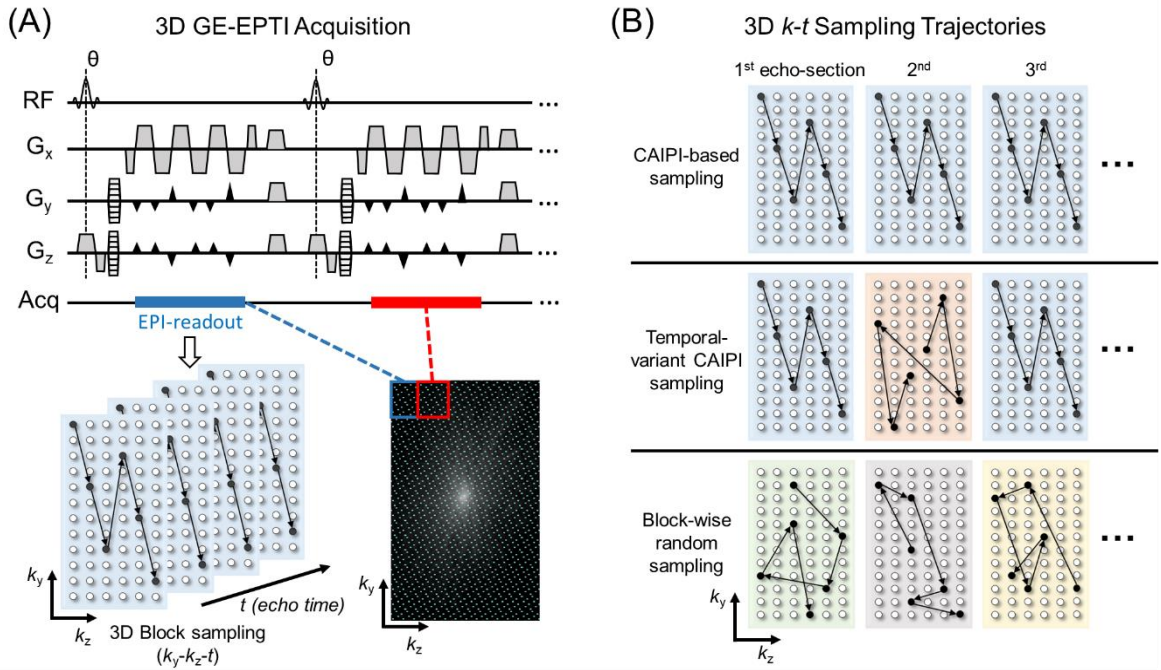

**FIG. 3.** The illustration of 3D GE-EPTI acquisition and three different designs of 3D spatiotemporal encoding. (A) The diagram of 3D GE-EPTI sequence and its block-wise sampling in $k_y$-$k_z$-$t$ space per EPTI-shot. Spatiotemporal CAIPI trajectory is employed in each EPTI readout section, and data blocks of different $k_y$-$k_z$ positions are acquired sequentially across TRs to fill the high-resolution $k_y$-$k_z$-$t$ space. (B) Three 3D spatiotemporal encodings at an acceleration factor of 72 (block size = 12×6): (i) CAIPI-based sampling, same CAIPI pattern of all echo-sections (ii) temporal-variant CAIPI sampling, shifted patterns between odd and even echo-sections, and (iii) block-wise random sampling.

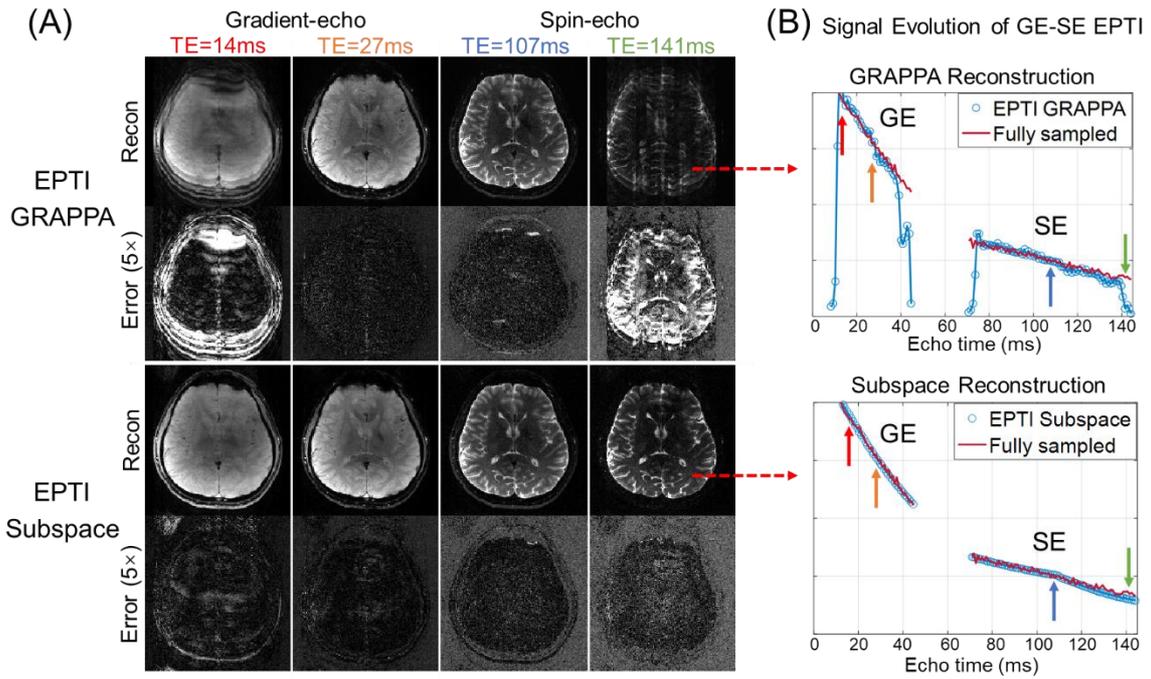

**FIG. 4.** Comparison of 2D GESE EPTI results reconstructed by $B_0$-informed GRAPPA and subspace reconstruction for a 9-shot 2D EPTI acquisition. Four images at different TEs (out of 120) are shown, where images at the early and late TEs show severe artifacts from the GRAPPA reconstruction, but the subspace method provided clean images. The signal drops of GRAPPA at the edge echoes are also shown in the signal curve. In contrast, subspace reconstruction obtained accurate signal evolution without signal drop or large fluctuation.

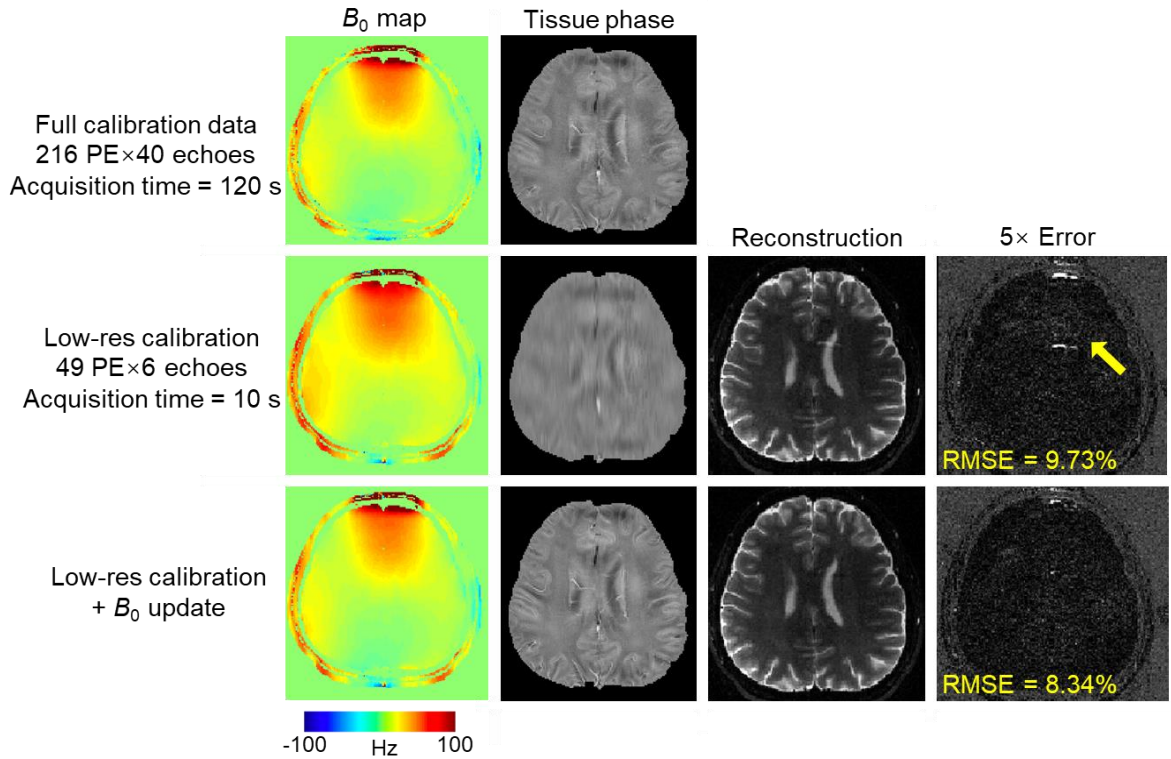

**FIG. 5.** The results of $B_0$ update evaluation. On the top, high-resolution $B_0$ map and tissue phase were calculated from the fully-sampled dataset with 216 PE and 40 echoes, which takes 120 seconds for 10 slices. In second row, reconstructed images and phase from a low-resolution calibration with 49 PE × 6 echoes are shown. Using low-resolution data, acquisition time can be reduced to only 10 seconds, but causes image artifacts in the reconstructed image. The bottom row shows the estimated high-resolution phase and image from the $B_0$ update algorithm, where the artifacts due to inaccurate phase are reduced.

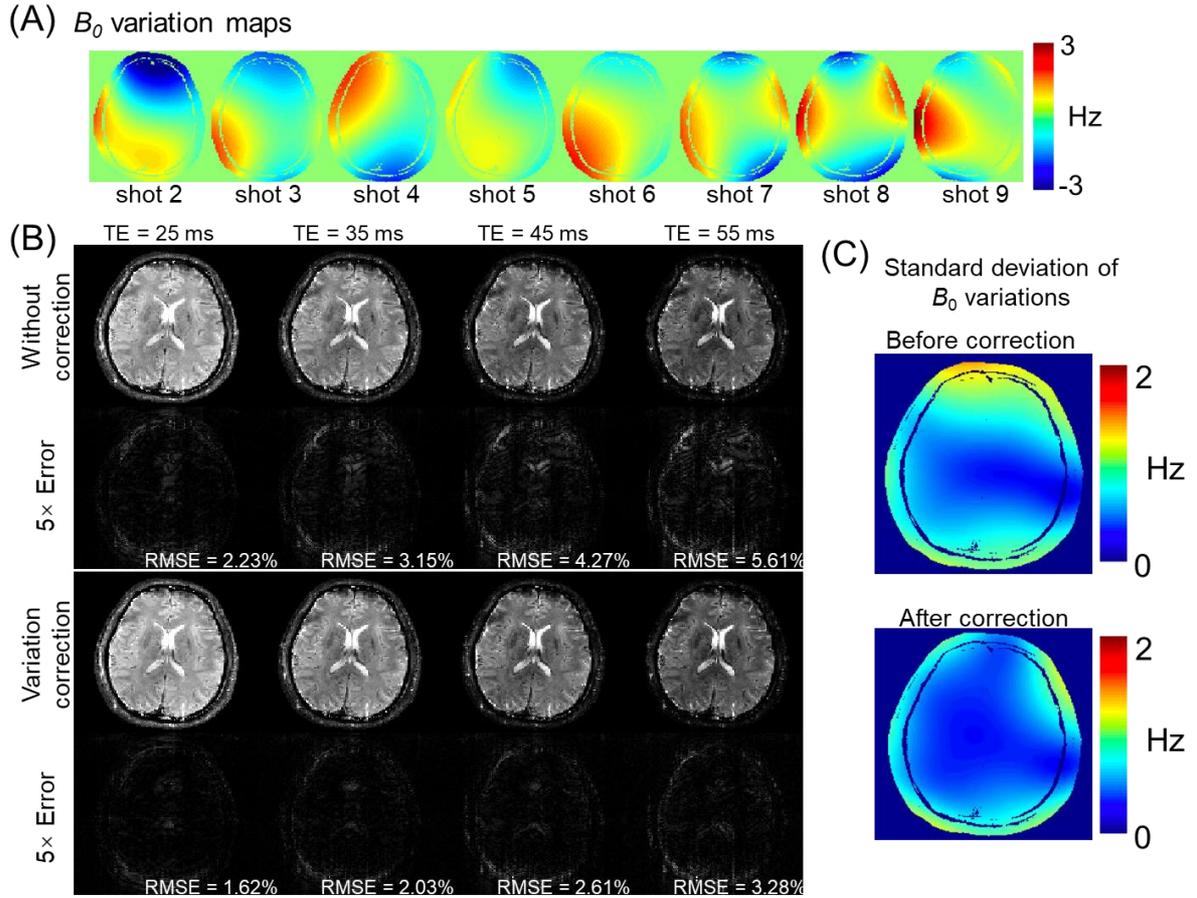

**FIG. 6.** Results of $B_0$ variation correction in the simulation test. (A) Realistic $B_0$ maps obtained from an in vivo scan across the 9 shots. (B) The reconstructed images and error maps (×5) with and without variation correction at four different TEs. (C) The standard deviation of $B_0$ variation across 9 shots before and after correction. After variation correction, the standard deviation and reconstruction errors are all reduced.

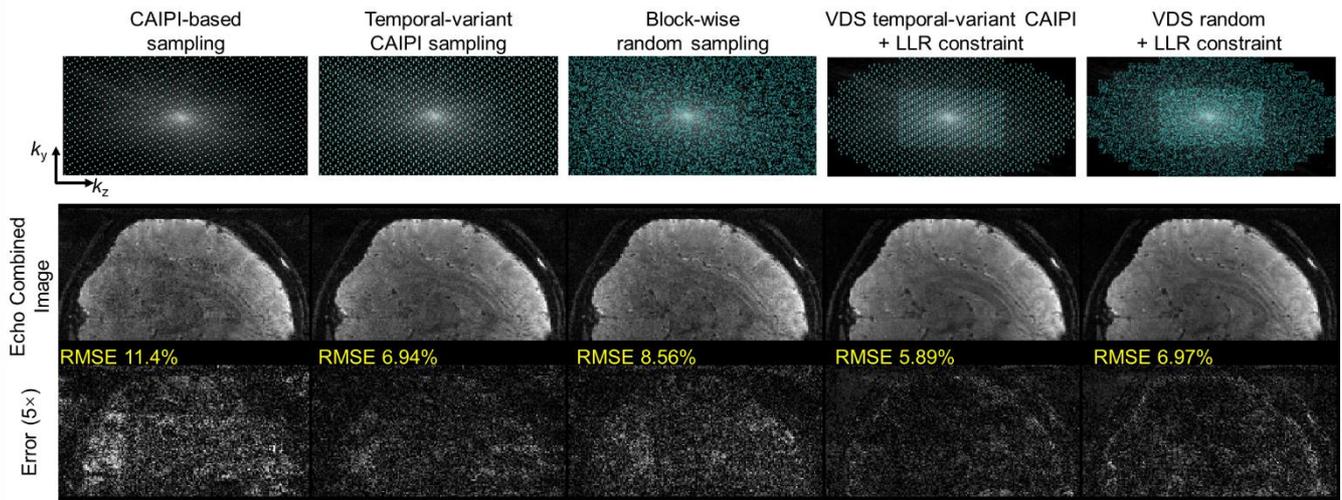

**FIG. 7.** Comparison of different encoding patterns in the retrospective undersampling experiment using 3D GE-EPTI dataset at 72× acceleration. The $k_y$-$k_z$ view of the five patterns that sums all the $k$-$t$ sampling over the $t$ dimension, reconstructed images, and 5× error maps are shown in the figure. CAIPI-based, temporal-variant CAIPI, block-wise random sampling are reconstructed by subspace approach without LLR constraint, and VDS temporal-variant CAIPI and VDS random sampling are reconstructed with LLR constraint. The temporal-variant CAIPI shows the lowest reconstruction errors in the non-VDS sampling, and its combination with VDS sampling and LLR further reduce the RMSE.

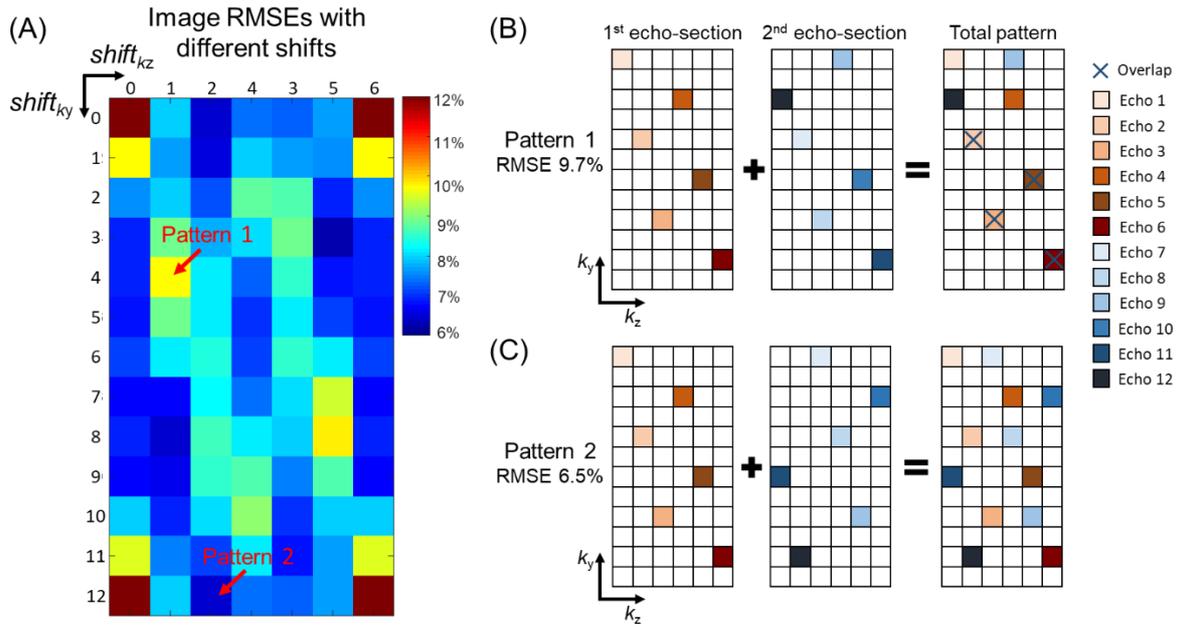

**FIG. 8.** (A) The image RMSEs of the temporal-variant CAIPI sampling with all different shifts between odd and even echo-sections are calculated. The block size is 12×6 ($k_y$×$k_z$), so the shift range of $k_y$ is 0-12 (0 and 12 are the same), the shift range of $k_z$ is 0-6 (0 and 6 are the same). Two different patterns are selected to analyze the factor that affects the performance at different shifts. The sampling pattern of odd echo-section, even echo-section and the total pattern are shown in Fig. 8B and C for Pattern 1 (4-1 shift) and Pattern 2 (12-0). Pattern 2 shows more complementary sampling in the $k_y$-$k_z$ block than Pattern 1, and achieved lower RMSE of the reconstructed image.

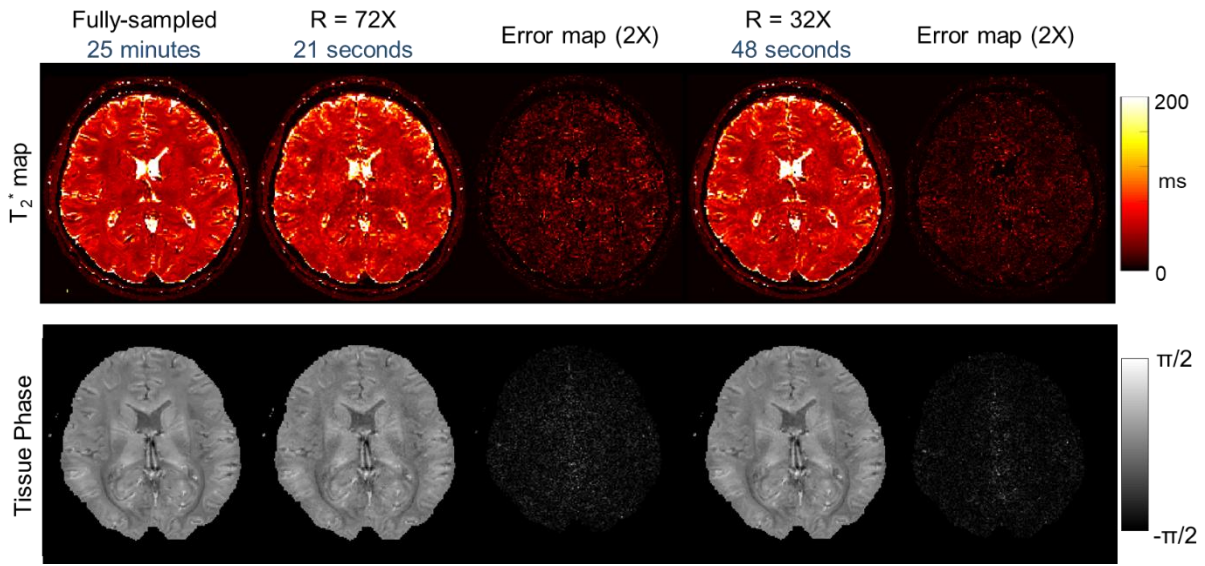

**FIG. 9.** The quantitative $T_2^*$ and tissue phase maps at 72× and 32× retrospective undersampling with VDS temporal-variant CAIPI. The scan time of fully-sampled scan was 25 minutes at 1.1 mm isotropic resolution, which can be reduced to 21 seconds and 48 seconds by 72× and 32× folds acceleration respectively. At R = 72, the tissue phase is comparable to reference, but $T_2^*$ map shows artifacts and noise. Using R = 32, the error of $T_2^*$ map is reduced with higher SNR, and the acquisition time is still less than 1 minute.

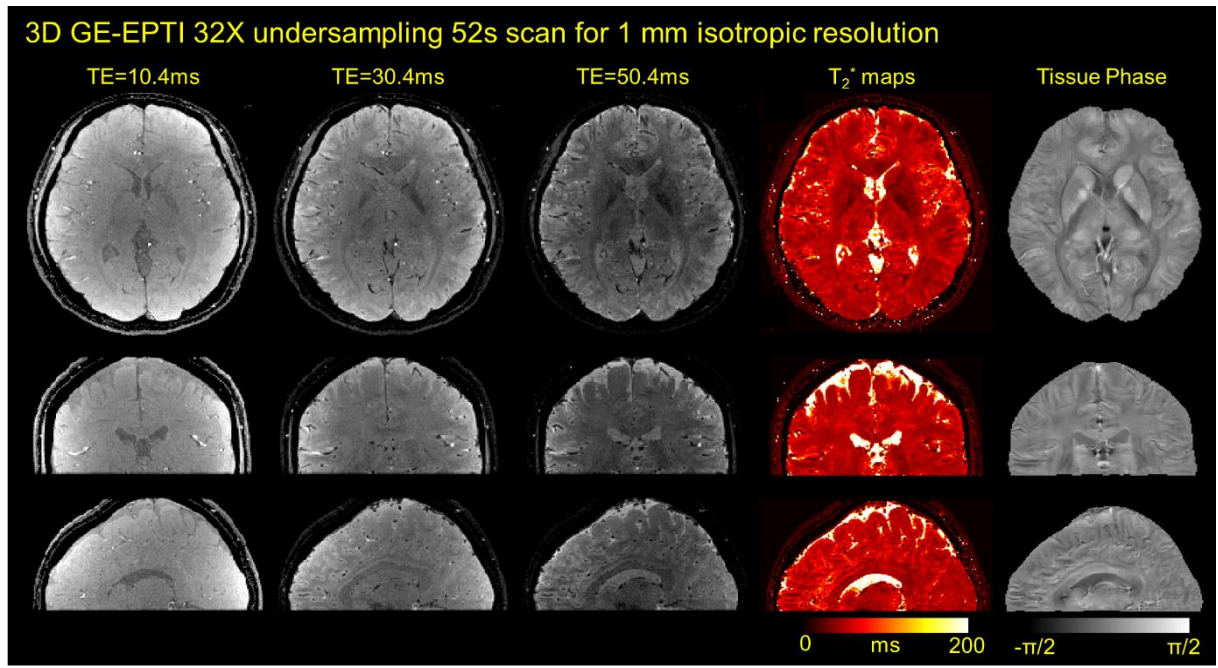

**FIG. 10.** The results of prospective undersampling experiment using 3D GE-EPTI with VDS temporal-variant sampling. Whole-brain $T_2^*$-weighted images with 57 echoes at 1 mm isotropic resolution was acquired in only 52 seconds. $T_2^*$-weighted images, quantitative $T_2^*$ maps and tissue phase are presented in three orthogonal views.

**Supporting Information**

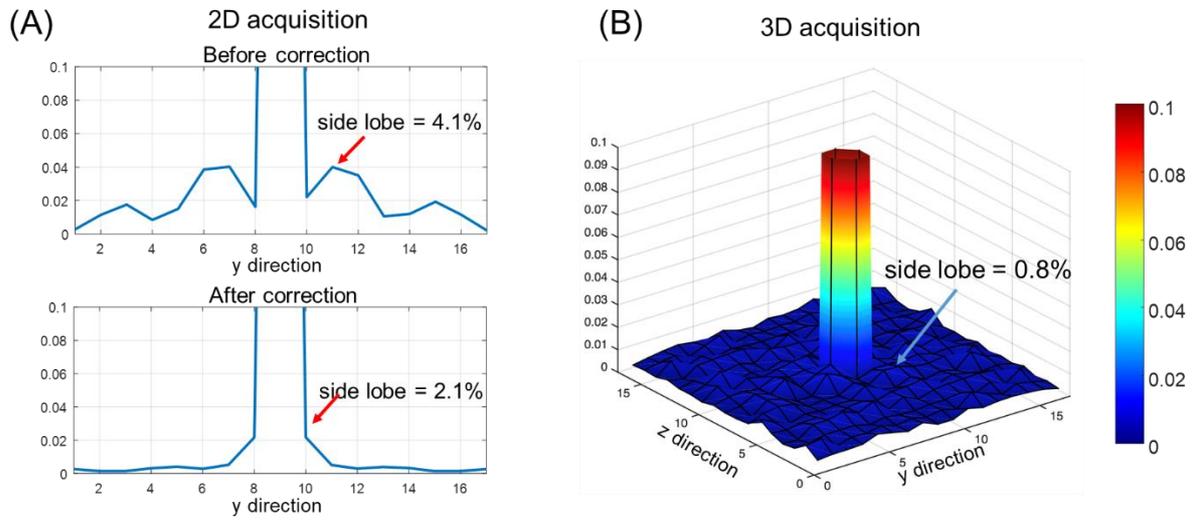

**FIG. S1.** (A) PSF analysis of 2D EPTI acquisition before and after $B_0$ variation correction. Before correction, $B_0$ variation with standard deviation of 1.5 Hz causes local smoothing with side lobe = 4.1%, and after correction, the side lobe is reduced to 2.1%. (B) PSF analysis for 3D EPTI acquisition with the same $B_0$ variation level (standard deviation = 1.5 Hz). In 3D acquisition, the largest side lobe is only 0.8%, because the effect of $B_0$ variation spreads along both $z$ and $y$ direction.